\RequirePackage{fix-cm}
\documentclass[twocolumn]{svjour3}          
\smartqed  
\usepackage{graphicx}
\usepackage[numbers]{natbib}
\usepackage{amsmath}

\usepackage [english]{babel}
\usepackage [autostyle, english = american]{csquotes}
\MakeOuterQuote{"}

\usepackage{soul}
\usepackage{url}
\usepackage[bookmarksopen,bookmarksdepth=2,breaklinks=true]{hyperref}
\usepackage[anythingbreaks]{breakurl}

\begin{document}

\title{Mode-space-compatible inelastic scattering in atomistic nonequilibrium Green's function implementations 
}


\author{Daniel A. Lemus\textsuperscript{1} (orcid.org/0000-0002-0759-8848)       \and
        James Charles\textsuperscript{1}        \and
        Tillmann Kubis\textsuperscript{1,2,3,4}
}


\institute{D. Lemus \at
              \email{dlemus@purdue.edu}           
              \and
              \textsuperscript{1} School of Electrical and Computer Engineering, Purdue University, 465 Northwestern Ave, West Lafayette, IN 47907
              \textsuperscript{2} Network for Computational Nanotechnology, Purdue University, 207 Martin Jischke Dr, West Lafayette, IN 47907, USA\\
              \textsuperscript{3} Purdue Center for Predictive Materials and Devices, Purdue University, West Lafayette, IN 47907, USA\\
              \textsuperscript{4} Purdue Institute of Inflammation, Immunology and Infectious Disease, Purdue University, West Lafayette, IN 47907, USA\\
}

\date{Received: date / Accepted: date}

\maketitle

\begin{abstract}
The nonequilibrium Green's function (NEGF) method is often used to predict transport in atomistically resolved nanodevices and yields an immense numerical load when inelastic scattering on phonons is included. 
To ease this load, this work extends the atomistic mode space approach of Ref.~\cite{Milnikov2012} to include inelastic scattering on optical and acoustic phonons in silicon nanowires. 
This work also includes the exact calculation of the real part of retarded scattering self-energies in the reduced basis representation using the Kramers-Kronig relations.
The inclusion of the Kramers-Kronig relation for the real part of the retarded scattering self-energy increases the impact of scattering.
Virtually perfect agreement with results of the original representation is achieved with matrix rank reductions of more than 97\%. 
Time-to-solution improvements of more than 200$\times$ and peak memory reductions of more than 7$\times$ are shown.
This allows for the solution of electron transport scattered on phonons in atomically resolved nanowires with cross-sections larger than 5~nm $\times$ 5~nm.

\keywords{Nanoelectronics \and NEGF \and Low-rank approximations \and Inelastic scattering \and Mode space}
\end{abstract}

\section{Introduction}
\label{sec:intro}
The characteristic length scale of state-of-the-art logic devices has reached dimensions with a countable number of atoms \cite{Charles2016,itrs}. At this scale, quantum effects such as tunneling, interference and confinement drastically change device performance \cite{pierret1996semiconductor,datta2005quantum,Ciraci2001,Jing}. Understanding and optimizing these effects almost always requires predictive models.
The nonequilibrium Green's function (NEGF) formalism is a well-accepted model for coherent and incoherent electron transport in nanodevices~\cite{Datta2000,Luisier2006}.

Characteristic nanoelectronic device dimensions contain a countable number of atoms, but a typical transistor contains hundreds to thousands of atoms in the volume of only a few cubic nanometers. Accurate basis representations such as the empirical tight binding method~\cite{boykin1991tight,Tan2015} usually contain tens of matrix elements per atom representing atomic orbitals \cite{luisier2006atomistic}. Solving the NEGF equations in a tight binding basis can be computationally cumbersome due to the required matrices consisting of thousands of rows and columns \cite{Bermeo2015,afzalian2015mode,guo2005nano}.
To ease this numerical load, the recursive Green's function method (RGF) \cite{Lake1997} provides a block-wise recursive solution for NEGF equations that can be discretized with block-tridiagonal sparse matrices \cite{Afzalian2017,chen2015approximate,kuzmin2013fast}. In that case, NEGF has been solved for nanodevices represented in realistic basis sets \cite{valencia2016grain,Wang2017,ilatikhameneh2015tunnel,chen2015nemo5}. With RGF, computational complexity depends on the cross-section and length of the device. In a typical nanowire device, the size of the blocks solved with the RGF method is directly proportional to the degrees of freedom $N$ in the cross-section of the device. Time-to-solution of matrix operations on these blocks scales on the order of $O(N^3)$. Memory scales on the order of $O(N^2)$.

The NEGF equations must be self-consistently solved with the Poisson equation that represents the electrostatic effects caused by the quantum mechanical evolution of the system \cite{Datta2000,datta2002non,trellakis1997iteration}. This introduces a degree of complexity to the solution of NEGF, since solving the equations is required multiple times.

An advantage of the NEGF and RGF methods is the ability to introduce incoherent scattering through self-energies, which represent device structure uncertainties such as roughness, alloy disorder and geometric errors, and temperature fluctuations through phonons~\cite{Datta2000, Kubis2011, he2012efficient, anantram2008modeling, Sarangapani2019, Zhang2010, Nguyen2013, Aldegunde2013, Valin2014, He2014, Ramayya2008}.
However, the introduction of incoherent scattering into the RGF solution introduces yet another degree of complexity through the self-consistent solution of retarded ($G^{R}$) and lesser ($G^{<}$) Green's functions. Their equations read symbolically
\begin{equation}
	G^{R}=(EI-H-\Sigma^{R})^{-1},
\end{equation}
\begin{equation}
    G^{<}=G^R\Sigma^{<}{G^R}^\dag,
\end{equation}
\noindent and the respective scattering self-energies
\begin{equation}
    \Sigma^{R}=G^{R}D^{R}+G^{R}D^{<}+G^{<}D^{R},
\end{equation}
\begin{equation}
	\Sigma^{<}=G^{<}D^{<}.
\end{equation}
\noindent In the above equations, $H$ is the electronic Hamiltonian, $I$ is an identity matrix, and $E$ is the electronic energy for which the Green's functions $G$ and self-energies $\Sigma$ are being solved. $D$ is the sum of environmental Green's functions with phonon, impurity and roughness information \cite{kubis2009theory,Kubis2011}. Within the self-consistent Born approximation the scattering self-energies and Green's functions are solved iteratively to achieve particle number conservation \cite{Kubis2011,kubis2009quantum,Luisier2010}.
It is worth mentioning that some alternatives to the self-consistent Born approximation of scattering exist, such as low-order approximations~\cite{Lee2017, Lee2016, Mera2012}, the B\"uttiker probe scattering model~\cite{Datta2000, Miao2016, Chu2019} and the multi-scale approach of Ref.~\cite{Geng2018}.
Although these methods are compatible with the mode space approach, they are beyond the scope of this work.

Many discretized degrees of freedom are common in atomistic representations, as well as the two layers of self-consistency, and usually result in heavy computational burdens. To ease this burden, incoherent scattering effects are often neglected in NEGF transport calculations \cite{datta2005quantum,venugopal2003simple,mamaluy2003efficient,rahman2003theory,shao2005nanoscale}. In the case of atomistic representations, even ballistic NEGF calculations often yield large computational loads. Such situations have motivated the introduction of a low rank approximation \cite{markovsky2011low} into NEGF \cite{chen2015approximate,mamaluy2003efficient,Zeng2013,hetmaniuk2015reduced,birner2009ballistic,he2012efficient}, which is often called the mode space approach \cite{afzalian2015mode,Afzalian2017,venugopal2003simple,Huang2017}.
Since scattering phenomena are important to retain in quantum transport simulations, the goal of this work is to introduce a low rank approximation that accurately retains scattering phenomena and is still based on an atomistic device representation.

\section{Method}
\label{sec:method}
\subsection{Mode space approach in tight binding}
Low-rank approximations such as the mode space method \cite{afzalian2015mode,Zeng2013,Huang2017} follow a common process: 
The system's Hamiltonian is transformed into a basis representation that allows for filtering of degrees of freedom that are unlikely to contribute to device operation. 
This reduces the rank of the system's Hamiltonian and thus the complexity of the NEGF equations. 
Choosing the eigenvectors of the Hamiltonian according to their eigen-energies often provides a good measure of filtering empty states \cite{Milnikov2012,birner2009ballistic}.
Unfortunately, this direct filtering fails in tight binding due to the appearance of spurious states \cite{Milnikov2012,Huang2017}. The method developed by Mil'nikov et al. \cite{Milnikov2012} removes these spurious states.

For completeness we repeat this method here: 
The first step of the method is to obtain the eigenvectors $\phi_i$ within the desired energy interval $\Delta\varepsilon$. The original basis Hamiltonian $H$ is transformed to a lower rank (mode space) basis using a rectangular transformation matrix $\Phi_{eig}$ constructed from the eigenvectors $\phi_i$:
\begin{equation}
 h=\Phi_{eig}^TH\Phi_{eig}.
\end{equation}
\noindent At this stage, the reduced Hamiltonian $h$ yields several unphysical states. A modified reduced Hamiltonian $\widetilde{h}$ is created by adding new orthogonal basis states $\widetilde{\Phi}$ ($\Phi_{eig}^{T}\widetilde{\Phi}=0$) such that  
\begin{equation}
 \widetilde{h}=
 \begin{vmatrix}
 h & X \\
 X^\dag & H_{\widetilde{\Phi}\widetilde{\Phi}}
 \end{vmatrix}
\end{equation}
\noindent where
\begin{equation}
 X=\Phi_{eig}^T H\widetilde{\Phi}.
\end{equation}
\noindent The added states $\widetilde{\Phi}$ do not deteriorate the basis and have no effect on non-spurious states.
The purpose of the added state $\widetilde{\Phi}$ is to remove the spurious states, thus $\widetilde{\Phi}$ are chosen such that they reduce the number of spurious states in the band structure. 
Since adding states to the basis keeps the physics unaltered \cite{Milnikov2012}, $\widetilde{\Phi}$ states are added until all spurious states within the energy interval $\Delta\varepsilon$ are removed and a transformation matrix $\Phi$ is produced.
The method by Mil'nikov et al. is therefore a minimization problem~\cite{Milnikov2012}.
Although devices in this work are homogeneous, and only require one basis to transform all portions of the nanowire, a heterogeneous device of varying cross-sections or materials may be transformed by use of multiple basis transformation matrices $\Phi$. It is therefore possible to introduce explicit defects such as roughness and impurities when $\Phi$ is obtained with such defects.

\subsection{Mode generation in NEMO5}
In this work, the mode space basis states are determined by following Mil'nikov et al.~\cite{Milnikov2012} with the ModeSpace solver \cite{Huang2017} of the multipurpose nanodevice simulation tool NEMO5 \cite{Steiger2011,nemo5}.
Details of this algorithm can be found in Refs.~\cite{Milnikov2012} and \cite{Huang2017}.
Ratios of the reduced $n$ and original $N$ Matrix ranks $n/N \leq 10\%$ are regularly achieved with this NEMO5 solver while the transport physics are preserved \cite{afzalian2015mode,Huang2017}. 
This has enabled speedups for ballistic NEGF simulations of up to 10,000 times \cite{Afzalian2017}.

\subsection{Expanding atomistic mode space to incoherent scattering simulations} \label{expanding_lra}
In this work, this method is augmented to handle incoherent scattering that allows for intermode transitions.
Scattering self-energies for the scattering of electrons on phonons are originally defined in a real space representation.
Electron scattering on acoustic and optical phonons via deformation potentials is considered following Ref.~\cite{Charles2016} (cf. Eqs.~1-6 of Ref.~\cite{Charles2016}).
Calculations in polar materials (such as InAs) include scattering of electrons on polar optical phonons as well~\cite{Riddoch1983}.
Since these self-energies are formulated in real space and require position information, an issue arises as this information is no longer directly available after a mode space basis transformation.

\subsection{Form factor transformation} \label{section_FF}
To make position information available for the solution of scattering self-energies while limiting the number of transformations, a form factor is introduced. This form factor is fully explained by Ref.~\cite{Afzalian2011}. The form factor $F$ contains all modes involved in the respective scattering process:
\begin{equation}
\label{form_factor}
 F_{i,j,k,l}=\sum_{\nu}\phi_i(\nu)\phi_j(\nu)\phi_k(\nu)\phi_l(\nu)
\end{equation}
\noindent where $i$,$j$,$k$,$l$ are indices of the $n$ real modes (columns) of the transformation matrix $\Phi$. The index $\nu$ is iterated through the $N$ rows of $\Phi$. 
We define each element of 
$\Sigma^{R,<}_{\text{acoustic}}$ and $\Sigma^{R,<}_{\text{optical}}$
of Eqs. 4-6 of Ref.~\cite{Charles2016} as $\Sigma_{i,j}$ and each element of a Green's function matrix 
$G^{R,<}$ as $G_{k,l}$.
We also define $C$ as the product of all scalar factors involved in each of the Eqs. 4-6 of Ref.~\cite{Charles2016}. The form factor elements $F_{i,j,k,l}$ are applied to the Green's function elements $G_{k,l}$ as follows:
\begin{equation}
\label{FF_selfenergy}
 \Sigma_{i,j}=\sum_{l}\sum_{k}CF_{i,j,k,l}G_{k,l}.
\end{equation}
\noindent In this way, all matrices remain in mode space.

\subsection{Approximation of form factor} \label{FF_approx}
The form factor $F$ is four-dimensional and scales rapidly with the number of modes in terms of memory ($O(n^4)$), time for construction ($O(n^4N)$) and time for application ($O(n^4)$). 
This can easily result in the form factor construction taking a significant amount of time and memory and application taking a significant amount of time.
Similarly to Ref.~\cite{Afzalian2011}, we have observed that eliminating off-diagonal elements of the form factor $F$, such that $F_{i,j,k,l}=0$ for $i \neq j$ and $k \neq l$, provides reasonable physical results.
This approximation corresponds to the lack of interaction between modes which are uncoupled in real space.
Therefore, no inter-mode scattering takes place when the form factor is diagonal.
This does not restrict inelastic scattering, since the electronic energy is a mode-independent parameter.
This approximation provides a memory-thin form factor with memory scaling on the order of $O(n^2)$. 
The construction complexity of the form factor is also reduced to $O(n^2N)$, while the application complexity is reduced to $O(n^2)$.
Note that although this yields an accurate calculation of self-energies $\Sigma^{R,<}_{\text{acoustic}}$ and $\Sigma^{R,<}_{\text{optical}}$, mode coupling terms ($G_{k,l}$ for $k \neq l$) must remain for an accurate calculation of electron density \cite{Afzalian2011}.

\subsection{Inclusion of real part of retarded scattering self-energies using Kramers-Kronig relations}
\label{subsec:method_KK}
The general form of the retarded scattering self-energy $\Sigma^R$ includes a principal value integral $\mathcal{P}$ of large  computational burden~\cite{Milnikov2012, Charles2016, Frey, Esposito2009,Svizhenko2003, Wacker2002}. 
$\Sigma^R(E)$ can be obtained by its separate real and imaginary parts~\cite{Frey,Esposito2009,Svizhenko2003} such that
\begin{equation} \label{eq:SigmaR_real}
    Re[\Sigma^R(E)] = \frac{i}{\pi}\mathcal{P}\int{dE'\frac{Im[\Sigma^R(E')]}{E-E'}.}
\end{equation}
\noindent Typically, the real part of the retarded self-energy is entirely excluded, and although the approximation often yields reasonable physical results~\cite{Milnikov2012,Svizhenko2003}, it is known that excluding the real part causes deviations.
In particular, off-state current densities are underestimated in this approximation~\cite{Charles2016, Frey, Esposito2009}.
Note that the real part of retarded self-energies shifts resonance energies and thus influences band edges and threshold voltages~\cite{kubis2009quantum}.
In this work, the exact real part of the retarded scattering self-energies is obtained using the Kramers-Kronig relations~\cite{Kronig1925}.
For each matrix element $\Sigma_{i,j}^R$ of a retarded self-energy, its real part $\Sigma(E)_{i,j,real}^R$ is obtained by applying the Kramers-Kronig relation on its imaginary part $\Sigma(E)_{i,j,imag}^R$. 
Using a Hilbert transform $\mathcal{H}$, the real part becomes:
\begin{equation} \label{eq:KK_simple}
    \Sigma(E)_{i,j,real}^R = \mathcal{H}(\Sigma(E)_{i,j,imag}^R).
\end{equation}
This Hilbert transform is performed using a fast Fourier transform (FFT), a multiplication in the Fourier space, and an inverse FFT afterwards~\cite{Cizek1970}.

\section{Results and discussion} \label{sec:results}
\subsection{Simulation setup}
\begin{figure}
  \includegraphics[width=0.48\textwidth]{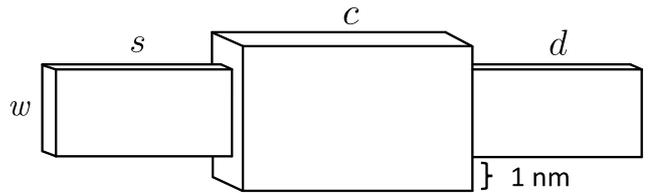}
  \caption{Schematic of the nanowire devices considered in this work with a \textit{w} $\times$ \textit{w} cross-section and a 1~nm gate oxide layer surrounding the center of the device. $s$ labels the source length, $c$ the channel length and $d$ the drain length of the device}
  \label{fig:1}
\end{figure}

To ensure the validity of the presented low-rank approximation for transport in nanowire devices including inelastic scattering, multiple tests were performed with NEMO5~\cite{Steiger2011,fonseca2013efficient,sellier2012nemo5}.
First, for validation, results of simulations in a mode space basis were benchmarked against calculations in the original tight binding basis. 
These result comparisons are shown in Sec.~\ref{section_validation}. 
Second, multiple performance tests comparing time-to-solution and peak memory improvements in mode space are shown in Sec.~\ref{section_performance} for various device widths \textit{w}.
The device used for both validation and performance tests was a \textit{w} $\times$ \textit{w} $\times$ 20.65~nm silicon nanowire as shown in Fig.~\ref{fig:1}, where \textit{w} is the variable width in nm of the square cross-section of the device.
The device had a 1~nm gate oxide layer surrounding the central region.
The original basis was a 10-band $sp^3d^5s^*$ tight binding model using the parameter set of Ref.~\cite{boykin2004valence}.
A source-drain bias of 0.2~V was applied to the device.
Note that the applied source-drain bias does not affect the validity of the presented method, and mode space calculations with higher source-drain voltages can be found in Refs.~\cite{Afzalian2017, Huang2017, Afzalian2018}.
The device was \textit{NIN} doped, with the $s = $~5.97~nm source and $d = $~6.66~nm drain regions having a $10^{20}$~cm\textsuperscript{-3} doping density and the central $c = $~8.02~nm intrinsic region having a $10^{15}$~cm\textsuperscript{-3} doping density.
The lengths $s$, $d$ and $c$ are labeled in Fig.~\ref{fig:1}.
Simulations of Si devices included both inelastic optical phonon and elastic acoustic phonon deformation potential scattering, applied to the NEGF equations through self-energies in the self-consistent Born approximation \cite{Charles2016,anantram2008modeling}. 
For polar materials, scattering on polar optical phonons was included as well.
The inhomogeneous energy grid was generated using an adaptive grid generator in NEMO5 \cite{Charles2016}.
Due to the high numerical load of the Kramers-Kronig relation for scattering in tight binding representations, the real parts of all scattering self-energies in the benchmarking scenarios of Secs.~\ref{section_validation}, \ref{section_performance} and \ref{simulating_beyond_existing} were neglected.
This is not the case in Sec.~\ref{subsec:KK_results}, where a non-zero real part of the scattering self-energy will be included.

An assessment of the real part of the retarded self-energies was done by comparison of the resulting current-voltage (I-V) characteristics shown in Sec.~\ref{subsec:KK_results}.
For this assessment, the material of the transistor in Fig.~\ref{fig:1} was chosen to be InAs, with two tested device widths $w =$ 2.42 nm and $w =   $ 3.63 nm.
Both devices had an $s = $ 5.97 nm p-type source doped at $5\times10^{19}$ cm\textsuperscript{-3}, an n-type $d = $ 9.66 nm drain doped at $2\times10^{19}$ cm\textsuperscript{-3} and a $c = $ 14.66 nm central undoped region.
A source-drain bias of 0.3~V was applied.
Since TFETs require the occupation of both electrons and holes, the method of Ref.~\cite{Milnikov2012} was applied to obtain modes for a wide energy window that included bands near the conduction and valence band edges.
The inclusion of holes also necessitates a proper definition of electrons and holes as states tunnel from valence band to conduction band in the TFET.
An interpolation method was applied as defined by Ref.~\cite{Charles2016} to avoid sharp transitions from holes to electrons or vice versa.
Simulations included optical phonon, acoustic phonon and polar-optical phonon scattering to represent the polar nature of InAs.
Due to the non-local nature of polar-optical phonon scattering, such a calculation would be very expensive even in a reduced basis.
To avoid this, a local scattering calculation was performed using a cross-section-dependent compensation factor defined in Ref.~\cite{Sarangapani2019}.
Compensating scaling factors of 30.0 and 26.56 were used in the calculation of polar-optical phonon scattering for the $w =~$2.42~nm and $w =~$3.63~nm devices respectively.
Note, the form factor approximation as described in Sec.~\ref{FF_approx} was not performed in this case.

\subsection{Validation of mode space simulation results} \label{section_validation}
\begin{figure}
  \includegraphics[width=0.48\textwidth]{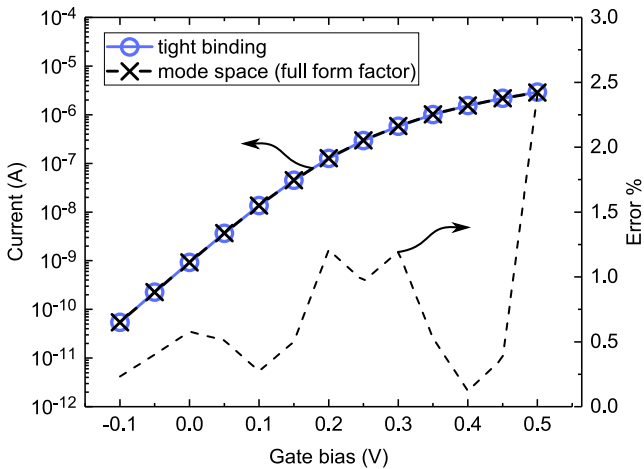}
  \caption{Current-gate-voltage (I-V) characteristic curve of a 3.26~nm $\times$ 3.26~nm $\times$ 20.65~nm silicon nanowire. The agreeing results prove the mode space approach provides a valid physical model. All simulations include inelastic scattering on phonons}
  \label{fig:2}
\end{figure}

\begin{figure}
  \includegraphics[width=0.48\textwidth]{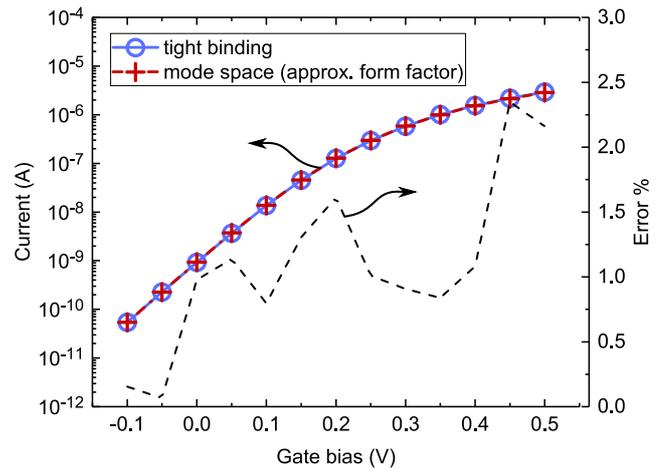}
  \caption{I-V curve of the 3.26~nm $\times$ 3.26~nm $\times$ 20.65~nm silicon nanowire of Fig.~\ref{fig:2} with an approximate form factor. The agreeing results justify the form factor approximation}
  \label{fig:3}
\end{figure}

\begin{figure}
  \centering
  \includegraphics[width=0.38\textwidth]{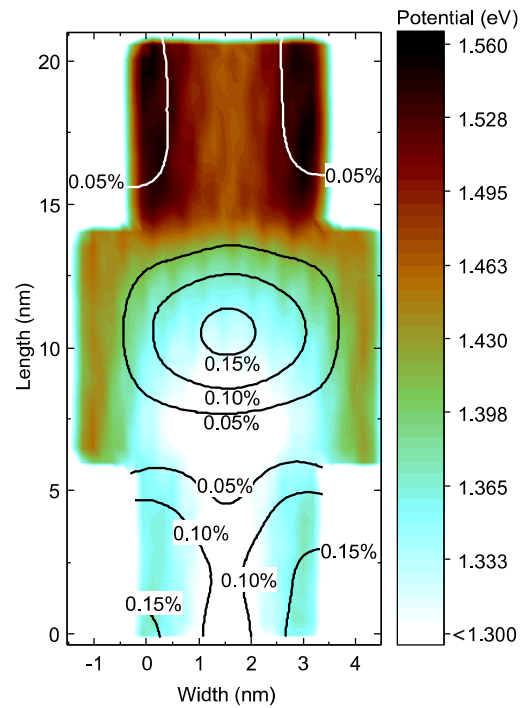}
  \caption{Potential profile (contour plot) of the center cross-section of the simulated 3.26~nm $\times$ 3.26~nm $\times$ 20.65~nm silicon nanowire device in original tight binding basis. Contour lines represent the relative error of the potential in mode space compared to tight binding representation}
  \label{fig:4}
\end{figure}

For validation, a silicon nanowire of width \textit{w} $= 3.26$~nm was used (see Fig.~\ref{fig:1}).
The mode space simulation had a reduction ratio $n/N$ of 2.8\%, transforming matrix blocks from 2880 $\times$ 2880 matrices to 81 $\times$ 81 matrices.
NEGF was solved using the scattering-compatible RGF algorithm \cite{anantram2008modeling}.
Fig.~\ref{fig:2} shows the current-voltage (I-V) characteristic curves of both the original tight binding basis and mode space basis for sweeping gate biases ranging from -0.1~V to 0.5~V.
The mode space scattering results of Fig.~\ref{fig:2} were obtained using the full form factor as described in Sec.~\ref{section_FF}. 
The virtually identical results of mode space and tight binding show that the mode space low-rank approximation provides a valid and highly efficient model for quantum transport simulations with inelastic scattering.
Fig.~\ref{fig:3} shows that the mode space approach with approximate form factors, as discussed in Sec.~\ref{FF_approx}, also yields results very close to those of the original basis calculations.
Fig.~\ref{fig:4} shows a contour plot of the potential profile of the center cross-section of the device for a tight binding simulation at the applied gate bias of 0.5~V. 
Contour lines show the relative error of the mode space potential profile results relative to the original tight binding data.
Note that the mode space method agrees with NEGF calculations in the original tight binding representation for many wire cross-sections as similarly well as those shown in Figs.~\ref{fig:2} and \ref{fig:3}. Similar benchmark data can be found in Refs.~\cite{Milnikov2012, Huang2017, Afzalian2018}.

\subsection{Assessment of computational performance} \label{section_performance}
\begin{figure}
  \includegraphics[width=0.48\textwidth]{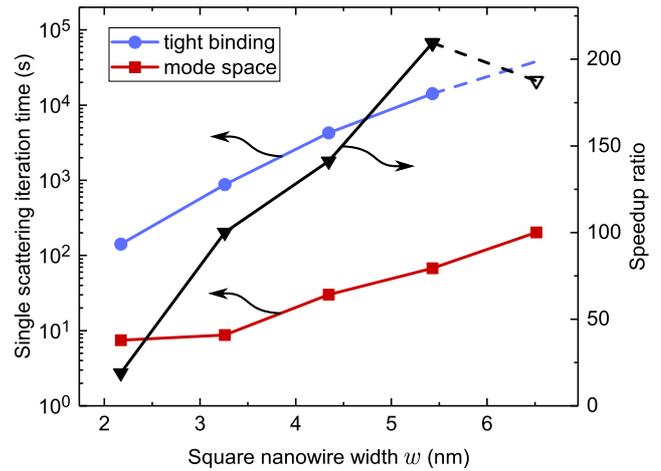}
  \caption{Time-to-solution for a single self-consistent Born iteration (left) and speedup ratio (right) with low-rank approximations for the 20.65~nm silicon nanowire of Fig.~\ref{fig:1} for various widths \textit{w}. The tight binding timing data was extrapolated beyond \textit{w}~$=$~5.43~nm using a power fitting function shown as a dashed line. All simulations include inelastic scattering}
  \label{fig:5}
\end{figure}

\begin{figure}
  \includegraphics[width=0.48\textwidth]{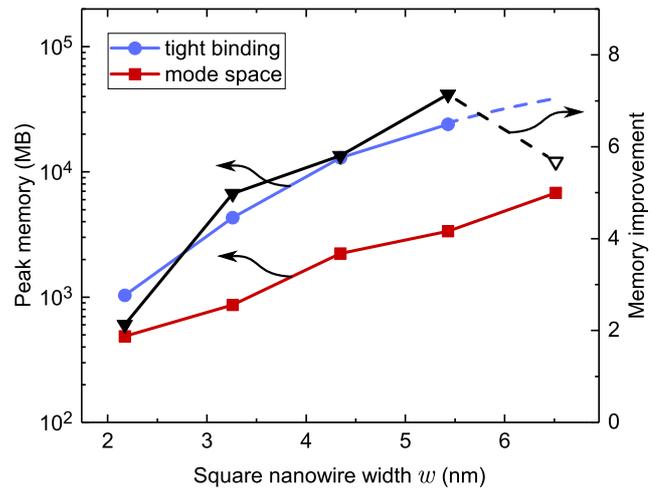}
  \caption{Peak memory (left) and memory improvement ratio (right) with low-rank approximations for 20.65~nm silicon nanowires of Fig.~\ref{fig:1} for various widths \textit{w}. All simulations include inelastic scattering}
  \label{fig:6}
\end{figure}

The device in Fig.~\ref{fig:1} was used with varying widths \textit{w} to measure performance improvements in NEMO5 time-to-solution and peak memory. 
Each width also had a corresponding mode space transformation matrix with its respective number of modes.
Correspondingly, the reduction ratios $n/N$ in Figs.~\ref{fig:5} and \ref{fig:6} vary.
The exact width values simulated were 4, 6, 8, 10 and 12 silicon unit cells and the respective reduction ratios $n/N$ were 5.6\%, 2.8\%, 2.9\%, 2.8\% and 3.0\%.
The lattice parameter of silicon was assumed to be 0.54~nm.
All performance simulations were performed with the same inputs of Sec.~\ref{section_validation}, with the exception that a fixed number of 256 energies was used.
Since results for the approximate form factor have been shown in Fig.~\ref{fig:3} to closely match those of the full form factor, mode space data for performance comparisons in this section were generated using the approximate form factor.
The Green's functions were solved for 256 energies with 1 energy per MPI process. 
Each MPI process was designated to a 32-core node on the Blue Waters petascale supercomputer at the University of Illinois at Urbana-Champaign \cite{bluewaters}.
Each MPI process was assigned 32 OpenMP threads for multithreaded matrix operations and form factor construction and application.
Fig.~\ref{fig:5} shows the average time (of 6 iterations) to compute a single self-consistent Born iteration. 
Each self-consistent Born iteration includes the time to compute the RGF algorithm as well as the time to compute lesser scattering self-energies $\Sigma^<$ and retarded scattering self-energies $\Sigma^R$ for optical and acoustic deformation potential inelastic scattering.
The calculation of scattering self-energies involves a large degree of communication between MPI processes as discussed in Ref.~\cite{Charles2016}.

The timing shown does not include the calculation of other aspects of quantum transport such as the solution of the Poisson's equation and the generation of the adaptive energy grid. 
This exclusion of such calculations can be justified by the fact that the time-to-solution is negligible when compared to the solution of NEGF.
In production runs, those calculations are performed only a small fraction of times when compared to the multiple self-consistent Born iterations per Poisson iteration.
The maximum speedup obtained with low-rank approximations for an iteration in this work was of 209.5 times. 
Due to computational limitations, the tight binding simulation for the point \textit{w}~$=$~6.52~nm was not assessed, since a single iteration would have taken about 38,000 seconds according to a power fitting function of the existing data. 
By extrapolating the data, the speedup for \textit{w}~$=$~6.52~nm is predicted to be of 187.5 times, as is shown in Fig.~\ref{fig:5}. It can be noted that this is lower than the speedup of \textit{w}~$=$~5.43~nm. 
This is likely due to the fact that the reduction ratio for \textit{w}~$=$~6.52~nm is slightly higher at 3.0\% than for \textit{w}~$=$~5.43~nm at 2.8\%.
Fig.~\ref{fig:6} shows the peak memory of the same simulations run for Fig.~\ref{fig:5}.
The maximum peak memory reduction was of 7.14$\times$.
Similarly to Fig.~\ref{fig:5}, a power fitting function was used to predict the peak memory for a device of \textit{w}~$=$~6.52~nm, which results in a predicted peak memory reduction of 5.67$\times$.


\subsection{Simulating beyond existing capabilities} \label{simulating_beyond_existing}
\begin{figure}
  \includegraphics[width=0.48\textwidth]{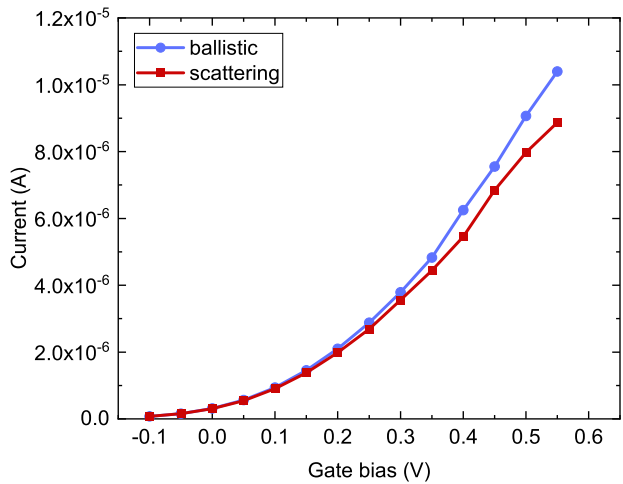}
  \caption{Comparison of I-V characteristics for a 5.43~nm $\times$ 5.43~nm $\times$ 20.65~nm n-type FET device for simulations with and without inelastic scattering. The reduction ratio $n/N$ for this simulation was 2.8\%. This device size significantly exceeds the largest nanowires possible to resolve in a scattered NEGF calculation in the original atomic representation}
  \label{fig:7}
\end{figure}

With the time-to-solution and memory footprint significantly reduced, the opportunity to simulate larger devices with complex physical phenomena such as incoherent scattering of multiple types (phonons, roughness, impurities) is now accessible. 
Ref.~\cite{Charles2016} describes the simulation of a circular nanowire, with acoustic and optical deformation potential scattering and a 10-band tight binding basis. 
The diameter of the cross-section of this device was 3~nm, and the device length was 27~nm.
Solution of an I-V characteristic curve took approximately 275 hours on 330 cores on the Blue Waters petascale supercomputer.
The peak memory was 60 GB per node, which is close to the maximum node memory of 64 GB. 
This device therefore approaches the limit of what can be simulated in a full basis representation such as tight binding.
To demonstrate the capability of solving larger devices in a reduced basis, a full I-V curve was generated for a square nanowire of Fig.~\ref{fig:1} with \textit{w}~$=$~5.43. 
Due to the different cross-sectional geometry this nanowire has over 4 times more atoms in the cross-section than the circular nanowire of Ref.~\cite{Charles2016}.
The reduction ratio $n/N$ for the square nanowire was of 2.8\%. 
Fig.~\ref{fig:7} shows an I-V characteristic curve for optical and acoustic phonon deformation potential scattering compared to that of a ballistic simulation. 
As expected, the on-current density is reduced by the inelastic scattering on phonons~\cite{Charles2016, kubis2009theory, Frey}.
The scattered transport simulation of the \textit{w}~=~5.43~nm device took approximately 160 total hours on 16,384 cores (2.62 million core hours) on the Blue Waters supercomputer.
We estimate that the same I-V calculation would take about 550 million core hours and 168 GB of memory in the original tight binding basis representation. 

\subsection{Assessment of real part of retarded self-energies}   \label{subsec:KK_results}

The 2-norms of the real and imaginary parts of the retarded self-energy $\Sigma^R$ can show the relative amplitude of their contributions.
Comparing the 2-norms of fully charge-self-consistent calculations is misleading, however, since scattering impacts the density of states:
The Poisson potential would compensate some of the density of state differences to accommodate the device's doping profile. 
Therefore, for this comparison only, scattering self-energies and Green's functions were solved self-consistently with a fixed Poisson potential.
That potential was deduced from a converged ballistic transport solution of the same device.
The calculations were performed for the on-state bias of 0.4~V.
Table~\ref{table:1} shows the 2-norm values of the real and imaginary parts of the $\Sigma^R$ when the Kramers-Kronig relation is observed and when the real part is set to 0. 
In both of the simulated cross-sections, the norm of the real part is comparable to the norm of the imaginary part.
Due to the reduced size of self-energy matrices, it was possible to perform the Hilbert transform on all energies without introducing memory issues and long computation times. Therefore in this work, all energy points generated by the adaptive energy grid in NEMO5~\cite{Charles2016} were included in the Hilbert transform of the self-energies.

\begin{figure}
    \centering
    \includegraphics[width=0.48\textwidth]{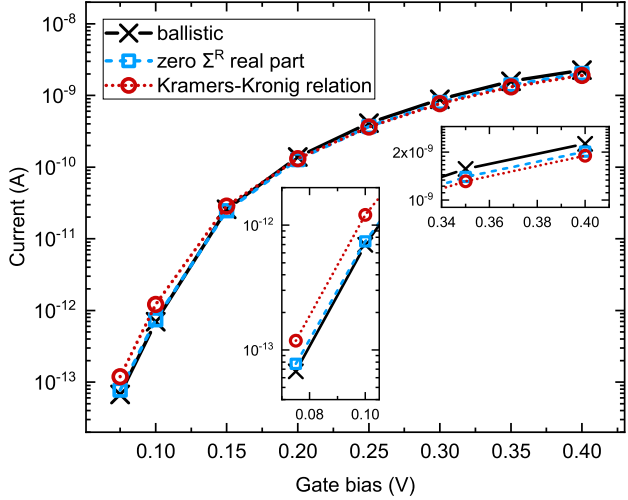}
    \caption{I-V characteristics for a 2.42~nm $\times$ 2.42~nm $\times$ 30.29~nm InAs TFET device solved in NEGF including incoherent scattering on polar optical phonons, acoustic phonons and optical deformation potential phonons. Scattering, even without a real part of $\Sigma^R$, increases the off-current densities and lowers on-current densities. When the real part of the retarded self-energy $\Sigma^R$ is included, the Kramers-Kronig relations are obeyed and scattering shows an even larger impact. The insets zoom into the first two and the last two points of the curves}
    \label{fig:8}
\end{figure}

\begin{figure}
    \centering
    \includegraphics[width=0.48\textwidth]{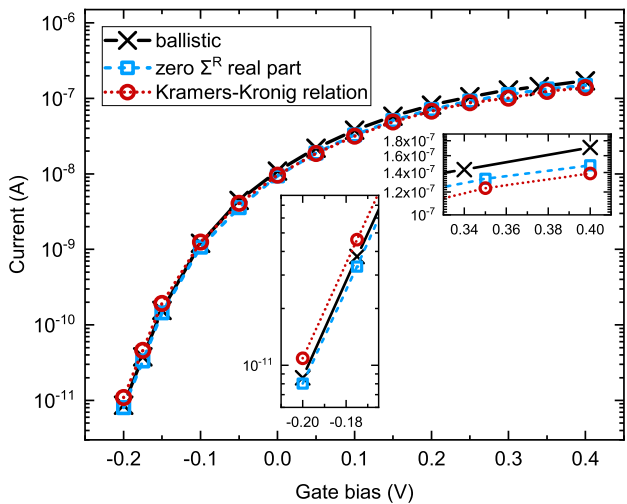}
    \caption{Similar to Fig.~\ref{fig:8}, I-V characteristics of a 3.64~nm $\times$ 3.64~nm $\times$ 30.29~nm InAs TFET device. The effects of scattering with and without a real part of $\Sigma^R$ are larger than in the smaller wire of Fig.~\ref{fig:8}}
    \label{fig:9}
\end{figure}

\begin{figure}
    \centering
    \includegraphics[width=0.48\textwidth]{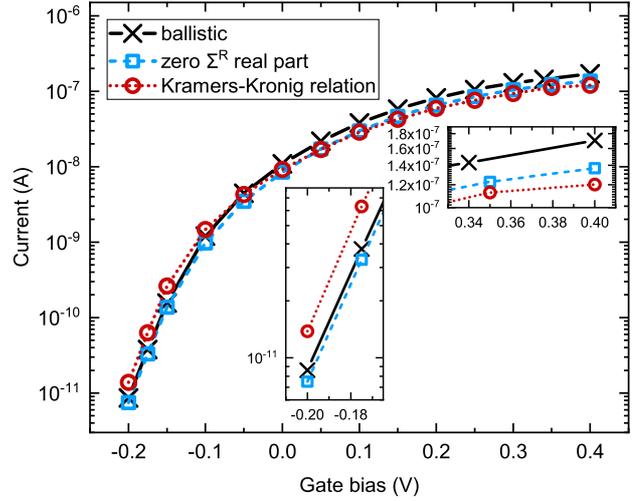}
    \caption{Similar to Fig.~\ref{fig:9}, I-V characteristics of a 3.64~nm $\times$ 3.64~nm $\times$ 30.29~nm InAs TFET device, but with scattering self-energies multiplied by 2}
    \label{fig:10}
\end{figure}

Figs.~\ref{fig:8} and \ref{fig:9} show the I-V characteristics of the \textit{w}~=~2.42~nm and \textit{w}~=~3.64~nm devices respectively. 
Both figures show the differences of the two scattering models (with and without the real part of $\Sigma^{R}$) when compared to the ballistic transport.
Incoherent scattering increases the off-current density due to scattering-supported gate leakage and decreases the on-current density due to stronger back-scattering. 
This is in agreement with findings in literature~\cite{Charles2016,Luisier2010,Frey,Svizhenko2003,Charles2018}.
It should also be noted that the real part of scattering self-energies has notable effects on devices of any dimension, e.g., 1D, 2D and 3D~\cite{Sarangapani2019, Frey}.
The impact of the real part of $\Sigma^R$ becomes more apparent in situations with larger scattering strengths, e.g. when higher temperatures, impurity scattering, or surface roughness scattering are present.
Fig.~\ref{fig:10} shows the I-V characteristics of the device in Fig.~\ref{fig:9} solved with NEGF when all electron-phonon scattering self-energies were multiplied by 2. 
More significant gate leakage and back-scattering effects can be observed than that shown in Fig.~\ref{fig:9}. 
More importantly, however, Fig.~\ref{fig:10} shows that the exact $\Sigma^R$ with a non-zero real part provides even higher scattering strengths than the approximate, zero real part case.
This impact is particularly visible if the charge self-consistency does not compensate deviations from the doping profile. 
This is exemplified in Fig.~1 of Supplementary Material 1 which shows the charge density in a transistor after a single scattering iteration both in mode space and full tight binding representations. This figure also confirms the physics of the real part of scattering is correctly covered in mode space.


\begin{table}[]
\begin{tabular}{|l||l|l|l|l|}
\hline
width \textit{w} (nm) & \multicolumn{2}{l|}{zero real $\Sigma^R$} & \multicolumn{2}{l|}{Kramers-Kronig} \\ \hline
 & real & imag. & real & imag. \\ \hline
2.42 & 0 & 0.1184 & 0.0965 & 0.1130 \\ \hline
3.64 & 0 & 0.1080 & 0.0920 & 0.1104 \\ \hline
\end{tabular}
\caption{2-norms of the retarded scattering self-energies $\Sigma^R$ solved in NEGF simulations of two InAs TFETs with a width \textit{w} and an applied gate bias of 0.4~V. The norm of the real part, calculated using the Kramers-Kronig relations, is comparable to the norm of the imaginary part, and must have a similar significance to simulation results}
\label{table:1}
\end{table}


\section{Conclusion}
In this work, the atomistic mode space approach of Ref.~\cite{Milnikov2012} has been augmented to handle inelastic scattering on various types of phonons. 
The method was verified and benchmarked against results solved in the original representation for silicon nanowires of various sizes.
Valid results were achieved with matrix ranks reduced down to 2.8\% of their original rank. 
Time-to-solution was improved by up to 209.5 times, and peak memory was improved by up to 7.14 times. 
A full I-V calculation was performed in mode space for a 5.43~nm $\times$ 5.43~nm $\times$ 20.65~nm silicon nanowire in a $sp^3d^5s^*$ tight binding basis, which represents a system size larger than can normally be atomically simulated including inelastic phonon scattering.
The solution of the real part of the retarded scattering self-energies $\Sigma^R$ with the Kramers-Kronig relations ensures the exact treatment of incoherent scattering.
It is demonstrated with calculations of various nanowires that the real part of $\Sigma^R$ contributes to transport similarly to the imaginary part. Therefore, a reliable prediction of transport in NEGF must solve for the total complex $\Sigma^R$.

\begin{acknowledgements}
The authors acknowledge financial support by Silvaco Inc.
This research is part of the Blue Waters sustained-petascale computing project, which is supported by the National Science Foundation (award number ACI 1238993) and the state of Illinois. Blue Waters is a joint effort of the University of Illinois at Urbana-Champaign and its National Center for Supercomputing Applications. This work is also part of the “Advanced Nanoelectronic Device Design with Atomistic Simulations” PRAC allocation support by the National Science Foundation (award number OCI-1640876).
The authors acknowledge the Texas Advanced Computing Center (TACC) at The University of Texas at Austin for providing Stampede2 resources that have contributed to the research results reported within this paper. URL: http://www.tacc.utexas.edu

\end{acknowledgements}

\bibliographystyle{unsrt}
\bibliography{references}


\end{document}